# On-chip Dicke-type magnon polaritons in the ultrastrong coupling regime via spatially separated nanomagnets

Shugo Yoshii[1,2†], Manuel Müller[3,4], Ryo Ohshima[1,2], Matthias Althammer[3,4],

Yuichiro Ando[1,2,5], Hans Huebl[3,4,6], and Masashi Shiraishi[1,2*]

[1]Department of Electronic Science and Engineering, Kyoto University, Nishikyo-ku, Kyoto 615-8510, Japan

[2]Center for Spintronics Research Network, Institute for Chemical Research, Kyoto University, Uji, Kyoto 611-0011, Japan

[3]Walther-Meißner-Institut, Bayerische Akademie der Wissenschaften, 85748 Garching, Germany

[4]Technical University of Munich, TUM School of Natural Sciences, Physics Department, 85748 Garching, Germany

[5]Department of Physics and Electronics, Osaka Metropolitan University, 1-1 Gakuencho, Sakai, Osaka 599-8531, Japan

[6]Munich Center for Quantum Science and Technology (MCQST), 80799 Munich, Germany

[†]Shugo Yoshii, Email address: yoshii.shugo.x63@kyoto-u.jp

[*]Masashi Shiraishi, Email address: shiraishi.masashi.4w@kyoto-u.ac.jp

**Quantum electrodynamics lies at the heart of hybrid quantum systems essential for future technologies. The thermodynamic limit of the Dicke model, a fundamental model describing these systems, predicts exotic quantum phenomena, such as equilibrium superradiant phase transitions and ground-state two-mode squeezing. However, the experimental realization of genuine Dicke systems has remained elusive due to the inevitable existence of gauge-invariant self-interaction terms that hinder such phenomena. Here, we report on the on-chip realization of a Dicke-type system utilizing ultrastrong magnetic-dipole interactions between collective excitations in a spatially separated ferromagnetic array and a superconducting resonator, resulting in creation of magnon polaritons. Crucially, this spatially separated architecture allows the cooperative enhancement of the coupling strength without increasing the self-interaction energy. We experimentally confirm the Bloch-Siegert shift, originating from the counter-**

**rotating terms, alongside the suppression of self-excitation terms required to observe critical Dicke physics. Our results establish a versatile platform, which provides the playground to explore quantum collective coupling physics and open pathways towards integrated quantum devices harnessing Dicke physics.**

## Main text

Quantum electrodynamics (QED) describing light–matter interactions is one of the most fundamental concepts in quantum science, underpinning technologies ranging from quantum information processing to quantum sensing[1–4]. An intriguing aspect of these interactions arises from anti-resonant processes, described by the simultaneous creation or annihilation of a photon and a material excitation – referred to as counter-rotating terms in the system's Hamiltonian. The Dicke Hamiltonian, which explicitly incorporates these counter-rotating terms for $N$-body two-level systems coupling to light, presents crucial opportunities both for deepening quantum science and for achieving noise-tolerant quantum technology. In the thermodynamic limit ($N \rightarrow \infty$), the Dicke model gives rise to profound quantum phenomena, including two-mode perfect squeezed ground states[5–7], virtual photon generation[8–10], and equilibrium superradiant phase transitions[11,12]. These phenomena form the foundation of next-generation quantum technologies, such as fast and robust quantum information processing and quantum computation[13]. These counter-rotating terms are so far neglected under typical strong coupling conditions, where the rotating-wave approximation remains valid. However, once coupling strengths surpass approximately 10% of the individual resonant frequencies, the system enters the ultrastrong coupling regime, rendering these previously negligible terms critically important. Thus, experimental platforms capable of reliably reaching the ultrastrong coupling regime have emerged as promising environments to realize Dicke model systems and foster the advancement of noise-tolerant quantum technologies.

Ultrastrong light–matter coupling has been demonstrated in diverse platforms, notably circuit QED using Josephson junctions[14–16] and semiconductor microcavities hosting Landau polaritons[17–19]. However, the experimental realization of the canonical Dicke model remains elusive. This is because, due to gauge invariance, the system in the thermodynamic limit corresponds to the Hopfield model with a self-interaction term, which enforces the no-go theorem forbidding equilibrium superradiant phase transitions[20,21]. Spin-based cavity QED systems initially appear to be ideal candidates for overcoming this limitation: the Zeeman coupling in magnetic-dipole interactions between cavity photons and collective spin excitations (magnons) naturally avoids the diamagnetic photon self-interaction ($A^2$ term) inherent to the minimal coupling. Although this distinct advantage of Zeeman coupling has been exploited in matter-matter interacting systems (e.g., magnon-spins coupling) to observe magnonic superradiant phase transition[22–25], its exploitation in circuit QED systems remains less explored. Furthermore, the gauge invariance of the Hamiltonian dictates that these systems suffer from a magnonic counterpart: the magnetic self-interaction of the magnons[26]. This magnetic term prevents the phase transition even in such spin-based systems[26]. Establishing a magnetic-dipole coupling system that addresses these challenges would significantly advance the exploration of Dicke physics in the circuit-based system, laying essential groundwork for noise-tolerant quantum technologies at the intersection of spintronics and quantum optics.

Here, we experimentally demonstrate ultrastrong magnon-photon coupling (magnon polaritons) with counter-rotating term dynamics, as evidenced by the Bloch-Siegert shift, and suppressed self-interaction terms. Our system (Fig. 2a) is a planar cavity-magnonic device integrating a ferromagnetic NiFe (permalloy) 30 nm thin film in the form of multiple isolated stripes, with a meander-type high-$T_C$ ($YBa_2Cu_3O_7$, i.e. YBCO) superconducting microwave resonator. Crucially, the formation of a magnon bright mode across the spatially separated ferromagnetic elements enables the cooperative enhancement of effective coupling strength while simultaneously avoiding the enhancement of magnon self-interaction term. This configuration realizes ultrastrong magnon polaritons with sizable counter-rotating terms but strongly suppressed self-interaction, representing a key achievement of on-chip Dicke-type magnon-polariton systems. Importantly, while operating in the macroscopic classical regime, the dynamics of our platform are mathematically isomorphic to the Dicke Hamiltonian in thermodynamic regime. Thus, this platform unlocks the potential for exploring critical Dicke physics, such as ground-state two-mode squeezing and the superradiant phase transition. Our approach establishes a scalable, lithographically-defined platform for studying exotic physical phenomena in the ultrastrong coupling regime and for merging spintronics with quantum optics, offering inviting routes to future device applications.

## Design principle for suppressing the magnon self-interaction terms

Although an individual spin weakly couples to a photon, the collective coupling $g \propto \sqrt{N/V_{\mathrm{p}}}$, where $N$ is the total number of spins and $V_{\mathrm{p}}$ is the photon volume, can be significantly enhanced through cooperative coupling (Dicke cooperativity) and by concentrating the microwave field using superconducting resonators. Strong photon confinement in superconducting resonators offers a promising route to realize large effective coupling strengths in magnon-photon interactions, owing to the reduced photon volume. The effective coupling strength between a magnon mode in a ferromagnet and the photon mode ($g_0$) is given by[27,28]:

$$g_0 = g_{\mathrm{s}}\sqrt{N}, \tag{1}$$

where $g_{\mathrm{s}}$ denotes the coupling strength between a single Bohr magneton and the photon, and $N$ is the total number of spins in the ferromagnet. $g_{\mathrm{s}}$ is proportional to $1/\sqrt{V_{\mathrm{p}}}$, on-chip superconducting resonators with diminished photon volumes can substantially enhance $g_{\mathrm{s}}$, enabling the observation of magnon polaritons even in thin ferromagnetic films with small magnetic volumes[29,30]. The uniform magnetization precession in a thin ferromagnetic film (the so-called Kittel magnon mode) interacts with the AC magnetic field of microwave photon resonator. For a sufficiently large magnetic moment, the interaction between the vacuum fluctuations of the microwave resonator and the magnetic modes becomes strong, exchanging energy in a coherent manner and forming hybrid magnon–photon states known as magnon polaritons.

To date, strategies to enhance the coupling strength between magnons and photons have primarily relied on increasing either the total number of spins or the vacuum fluctuation amplitude of the photon mode[31,32]. However, collective enhancement of the effective coupling strength inevitably induces a simultaneous enhancement of the self-interaction term, as predicted by the Hopfield model, a general describing coupled bosonic system (Fig. 1a). This scaling leads to the constraints imposed by the no-go theorem, which forbids the equilibrium superradiant phase transition[26].

To overcome this fundamental limitation, we introduce a distinct architecture comprising a single uniform photonic mode coupled to multiple spatially separated collective magnon modes in thin ferromagnetic films. In this scheme, each individual magnon mode couples to the photon mode with a single-element coupling strength $g_0$ (see Fig. 1b). The interaction between the photon mode and the collective magnon mode is described by the gauge-invariant Hopfield Hamiltonian, incorporating the self-interaction terms[26,32]. Considering the collective bright magnon mode formed from these individual modes via the photon-mediated coupling (see Methods for the derivation), the effective Hamiltonian is expressed as:

$$\begin{aligned}\mathcal{H} = \hbar\omega_{\mathrm{p}}\hat{a}^{\dagger}\hat{a} + \hbar\omega_{\mathrm{m}}\widehat{M}_{\mathrm{B}}^{\dagger}\widehat{M}_{\mathrm{B}} + \hbar G'_{\mathrm{eff}}\left(\hat{a}^{\dagger}\widehat{M}_{\mathrm{B}} + \hat{a}\widehat{M}_{\mathrm{B}}^{\dagger} + \hat{a}^{\dagger}\widehat{M}_{\mathrm{B}}^{\dagger} + \hat{a}\widehat{M}_{\mathrm{B}}\right) \\ + \hbar D_{\mathrm{m}}\left(\widehat{M}_{\mathrm{B}}^{\dagger} + \widehat{M}_{\mathrm{B}}\right)^{2} + \hbar D(\hat{a}^{\dagger} + \hat{a})^{2},\end{aligned} \qquad (2)$$

where, $\hat{a}$ ($\hat{a}^{\dagger}$) is the annihilation (creation) operator of photons, $\widehat{M}_{\mathrm{B}}$ ($M_{\mathrm{B}}^{\dagger}$) is annihilation (creation) operator of a collective magnon bright mode, $\omega_{\mathrm{p}}$ denotes the uncoupled photon resonance frequency, $\omega_{\mathrm{m}} = \gamma\mu_0\sqrt{H(H + M_{\mathrm{eff}})}$ is the uncoupled Kittel mode frequency dependent on the external magnetic field $H$. Here, $\gamma$ is the gyromagnetic ratio, $\mu_0$ is the vacuum permeability, and $M_{\mathrm{eff}}$ is the effective magnetization. $\widehat{M}_B$ is the operator for the collective magnon bright mode. To account for the dispersive nature of the magnetic coupling in the ultrastrong regime, the effective coupling strength is defined as $G'_{\mathrm{eff}} = G_{\mathrm{eff}}\sqrt{\frac{\omega_{\mathrm{m}}}{\omega_{\mathrm{p}}}}$. In addition, $G_{\mathrm{eff}} = g_0\sqrt{n}$, where $n$ is the number of the permalloy elements. Notably, the bright magnon mode exhibits a cooperative enhancement of the coupling strength, characterized by a $\sqrt{n}$ scaling, which equally affects both rotating and counter-rotating terms ($\hat{a}^{\dagger}\widehat{M}_{\mathrm{B}}^{\dagger} + \hat{a}\widehat{M}_{\mathrm{B}}$).

Importantly, this spatial separation strategy renders the magnon self-interaction term independent of the number of ferromagnetic elements: $D_{\mathrm{m}} = \frac{{g_0}^2}{\omega_{\mathrm{p}}}$ (see details in Methods) because the short-range nature of dipolar interactions prevents cross-coupling between separated ferromagnetic elements. Thus, unlike the effective coupling strength of magnon bright mode with photons, the magnon self-interaction term arises from the local confinement of magnons within each ferromagnetic element. Consequently, our non-local coherent coupling scheme decouples the cooperative enhancement of the effective coupling strength from the enhancement of the self-interactions of spins.

### Device implementation and spectral analysis

To realize this, we fabricate a lumped-element superconducting resonator from the high-$T_c$ superconductor, YBCO, incorporating a meander-shaped inductor (Fig. 2a) with permalloy stripes. This design supports a strongly confined and uniform AC microwave magnetic field on the inductor surface (see Methods) and achieves a quality factor of approximately 3600 (Fig. 2b). YBCO is a type-II superconductor and thus shows high resilience against external magnetic fields and principally operates up to the critical temperature of 90 K (see Supplemental Information). This enables the clear observation of magnon-photon hybridization for applied magnetic fields of 150 mT and moderate temperatures of 10 K.

The patterned 30 nm-thick permalloy stripes in these hybrid systems host collective magnon modes that couple via Zeeman interaction to the photon mode in a YBCO resonator. Figure 2c presents the transmission spectra normalized to a reference traces ($|\Delta S_{21}|$) taken at $\mu_0 H = 30$ mT for the resonator with 26 permalloy stripes in a YBCO chip (YBCO#1) as a function of the external positive magnetic field, which allows for the formation of magnon bright mode via photon-mediated coupling. Figures 2d-f show the enlarged spectra for 0, 7, and 14 mT with the fitting result of Eq. (5), respectively. Note that the larger linewidth of the spectra can be attributed to the larger magnon damping than photon damping (see the details in Methods). These spectra exhibit characteristic field dependence of the microwave resonance frequency, anticrossing spectra between magnon and photon modes, which is the hallmark of magnon-photon hybridization.

Within a quantum mechanical formalism, the magnon–polariton eigenfrequencies $\omega_{\mathrm{mp}}^{\pm}$ are determined by the following quartic equation (see Methods for the derivation):

$$\left({\omega_{\mathrm{mp}}^{\pm}}^2 - {\omega_{\mathrm{p}}}^2 - 4D\omega_{\mathrm{p}}\right)\left({\omega_{\mathrm{mp}}^{\pm}}^2 - {\omega_{\mathrm{m}}}^2 - 4D_{\mathrm{m}}\omega_m\right) - 4{\omega_{\mathrm{m}}}^2{G_{\mathrm{eff}}}^2 = 0. \quad (3)$$

To systematically evaluate the self-interaction energy relative to the fundamental limits imposed by gauge invariance, we parameterize the coefficients as $D = \alpha/\omega_{\mathrm{m}}$ and $D_{\mathrm{m}} = \beta/\omega_{\mathrm{p}}$. Here, $\alpha$ and $\beta$ serve as scaling parameters for the photon and magnon self-excitation terms, respectively. This parameterization facilitates a direct comparison with the theoretical scaling of self-interaction terms ($D, D_{\mathrm{m}} \propto {G_{\mathrm{eff}}}^2/\omega$). Extracting these parameters via spectral fitting allows us to quantitatively verify the condition for achieving a soft mode where the energy of the lowest polariton branch vanishes – a spectrum signature of superradiant phase transition.

Equation (3) serves as the basis for fitting the experimental spectra and extracting the coupling parameters. Figure 2g presents the fitting result with the extracted coupling parameters: $\omega_{\mathrm{p}}/2\pi$ = 5.041 GHz, $\mu_0 M_{\mathrm{eff}}$ = 1.108 T, and $G_{\mathrm{eff}}/2\pi$ = 512.7 MHz. These parameters yield a $G_{\mathrm{eff}}/\omega_{\mathrm{p}}$ ratio of 0.102, exceeding the widely accepted threshold (0.1) that defines the ultrastrong coupling regime[13]. The presence of counter-rotating terms modifies the polariton energy spectrum, manifesting as the Bloch-Siegert shift ($\Delta f_{\mathrm{BS}}$)[33]. This shift provides direct experimental evidence for counter-rotating terms in strongly coupled systems and can be identified by deviations from the dispersion curves obtained through quantum mechanical framework[15,34,35]. To elucidate this effect, we simulated magnon-polariton spectra based on the effective Hamiltonian shown as Eq. (2), explicitly comparing cases with and without counter-rotating terms by using the identical coupling parameters ($G_{\mathrm{eff}}, \omega_{\mathrm{p}}$ and $\omega_m$) obtained from the full Hamiltonian fit (Figs. 2h and i; see Methods for further details on dispersion calculations). The difference between two simulated spectra directly quantifies $\Delta f_{\mathrm{BS}}$ up to 60 MHz, aligning closely with our experimental observations and conclusively demonstrating the presence of counter-rotating terms.

We also quantitatively evaluated the photon and magnon self-interaction terms through the spectra fitting (Fig. 2g). The extracted scaling parameters are $\alpha = (4.96 \pm 0.34) \times 10^{17}$ (rad/s)$^2$ for photons and $\beta = (3.94 \pm 3.86) \times 10^{17}$ (rad/s)$^2$ for magnons. These values correspond to the self-interaction energy of $D$ = 2.50 MHz and $D_{\mathrm{m}}$ = 1.98 MHz, respectively, at the vacuum splitting condition $\omega_{\mathrm{p}} = \omega_{\mathrm{m}}$. To assess the suppression of the scaling law in the photon and magnon self-interaction terms, we normalized these coefficients by the squared effective coupling strength. We obtained suppression factors of $\xi_{\mathrm{p}} = \alpha/{G_{\mathrm{eff}}}^2 \sim 0.048$ for photons and $\xi_{\mathrm{m}} = \beta/{G_{\mathrm{eff}}}^2 \sim 0.038$ for magnons. These small values confirm a significant suppression of both self-interaction terms in our system. The suppression of the photon term ($\xi_{\mathrm{p}}$) is naturally expected in spin-based cavity systems[22–25], where the diamagnetic $A^2$ interaction is minimal compared to electric-dipole systems. More notably, the suppression of the magnon self-interaction term ($\xi_{\mathrm{m}}$) is directly attributed to the spatial separation of the ferromagnetic elements. Crucially, our spatially separated architecture physically enables the original premise of the Dicke supperadiance model: an ensemble of macroscopic emitters with negligible direct dipole-dipole interactions[36]. Because the ferromagnetic stripes are spatially isolated, inter-element static magnetic crosstalk is inherently eliminated (see Supplemental Information No. 5). Consequently, the total self-interaction energy of the system does not scale collectively.

This physical boundary condition – rather than an arbitrary mathematical truncation – is the exact mechanism that dictates the structural suppression of the effective self-interaction terms, breaking the standard scaling tied to the no-go theorem in continuous bulk media. By bypassing the long-range interactions, our architecture circumvents the standard scaling tied to the no-go theorem in continuous bulk media, where the condition ($\xi_{\mathrm{m}} > 1$) cannot be avoided[26]. This architecture decouples the cooperative enhancement of the effective coupling strength ($\propto \sqrt{n}$) from the enhancement of the self-interaction term, as derived in the Methods section. Remarkably, the experimentally determined magnon suppression factor ($\xi_{\mathrm{m}} \approx 0.038 \pm\ 0.037$) is consistent with the theoretical prediction based on the number of elements ($n = 26$): $\xi_{\mathrm{theory}} \approx 1/n \approx 0.038$. This consistency validates our theoretical model, where $D_{\mathrm{m}} = \frac{{g_0}^2}{\omega_{\mathrm{p}}}$ and $G_{\mathrm{eff}} = g_0\sqrt{n}$, demonstrating that the spatial separation scheme successfully suppresses the self-interaction term by harnessing the magnon bright mode in the separated ferromagnetic elements.

Moreover, the critical coupling strength required for achieving zero polariton eigenfrequency is determined by:

$$ {G'_{\mathrm{eff,c}}}^2 = \frac{{G'^{0}_{\mathrm{eff,c}}}^2}{2\xi_{\mathrm{p}}\xi_{\mathrm{m}}}\left\{(1-\xi_{\mathrm{p}}-\xi_{\mathrm{m}}) - \sqrt{(1-\xi_{\mathrm{p}}-\xi_{\mathrm{m}})^2 - 4\xi_{\mathrm{p}}\xi_{\mathrm{m}}}\right\}, \tag{4} $$

where $G'^{0}_{\mathrm{eff,c}} = \sqrt{\omega_{\mathrm{p}}\omega_{\mathrm{m}}}/2$ is the critical coupling strength for the Dicke Hamiltonian[6,21]. For our estimated values $\xi_{\mathrm{p}} = 0.048$ and $\xi_{\mathrm{m}} = 0.038$, $\xi_{\mathrm{p}} + \xi_{\mathrm{m}} < 1$ is satisfied, ensuring a finite critical coupling strength for the phase transition. This situation is apart from the ultrastrong coupling system which suffers from the enhancement of self-interaction terms aligning with the effective coupling strength[38]. The calculated critical coupling strength $G_{\mathrm{eff,c}}/2\pi$ is approximately 2.64 GHz, nearly five times greater than the effective coupling strength experimentally estimated as shown in Fig. 2g.

Figures 2h and 2i further illustrate simulated spectra derived from the ideal Dicke Hamiltonian – representing the limit of suppressed self-interactions – showing excellent agreement with experimental spectra at magnetic fields away from zero. These findings strongly suggest that around the vacuum splitting condition, magnon polaritons in the ultrastrong coupling regime effectively conform to the Dicke-like Hamiltonian, where counter-rotating terms responsible for squeezed magnon-photon states remain significant. In subsequent discussions, the Dicke model Hamiltonian will be utilized to analyze the lower polariton branch of the observed frequencies to precisely estimate coupling parameters.

## Dicke cooperativity in the coupling of one photon mode with separated magnon modes

We experimentally verify the cooperative nature of the coupling which underpins our theoretical model. Following Eq. (2), the effective coupling strength $G_{\mathrm{eff}}$ scales proportionally to the square root of the number $n$ of permalloy stripes coherently coupled to the photon mode, leading to cooperative enhancement. To verify this scaling behaviour experimentally, we prepared two YBCO chips (labelled YBCO#1 and YBCO#2), each incorporating resonators patterned with varying numbers of permalloy stripes ($n$ = 1, 8, 16, 26 for YBCO#1, and $n$ = 1, 26 for YBCO#2). The normalized transmission spectra of YBCO#1 chip, displayed in Figs. 3a-d, reveal distinct magnetic field-dependent behaviour characteristic of magnon-photon coupling. By fitting the lower polariton branch of each spectrum using the Dicke model, we extracted the $n$-dependence of $G_{\mathrm{eff}}$. Given that the single-Bohr-magneton coupling $g_{\mathrm{s}}$ is proportional to the square root of the uncoupled photon resonance frequency, $f_{\mathrm{p}}$[28], we introduce a normalized coupling strength $\varepsilon = G_{\mathrm{eff}}/2\pi\sqrt{f_{\mathrm{p}}}$, which should scale as $\sqrt{n}$. Figure 3e indeed confirms this linear relationship, with fitted slopes $\mathcal{S} = 1520\sqrt{\mathrm{Hz}}$ for YBCO#1 and $\mathcal{S} = 1791\sqrt{\mathrm{Hz}}$ for YBCO#2. This clearly demonstrates that the observed enhancement in $G_{\mathrm{eff}}$ originates from the formation of a collective magnon bright mode, consistent with Dicke cooperativity. The observed variation in the coupling coefficient $\mathcal{S}$ between YBCO#1 and YBCO#2 likely arises from differences in single-Bohr-magneton coupling $g_{\mathrm{s}}$, since $\mathcal{S} = g_0/\sqrt{f_{\mathrm{p}}} \propto g_{\mathrm{s}}$, where $g_{\mathrm{s}}$(YBCO#1) $\approx$ 28.4 Hz and $g_{\mathrm{s}}$(YBCO#2) $\approx$ 35.8 Hz.

This discrepancy may be attributed to variations in surface roughness between the YBCO chips (see Supplementary Information No. 3).

## $G_{\mathrm{eff}}$-dependence of Bloch-Siegert shift

According to the full Hamiltonian (Eq. (2)) of coupling system, the counter-rotating terms ($\hat{a}^{\dagger}M_{\mathrm{B}}^{\dagger} + \hat{a}\widehat{M}_{\mathrm{B}}$) induce a Bloch–Siegert shift—a characteristic frequency shift in the magnon-polariton modes[15,33,34], whose magnitude is controllable by altering the number of participating ferromagnetic elements. Demonstrating a Dicke-cooperative enhancement of this Bloch–Siegert shift constitutes compelling evidence for the presence and significance of counter-rotating interactions between magnons and photons. Figure 4a illustrates the $\Delta f_{\mathrm{BS}}$ as a function of external magnetic fields for resonators comprising varying numbers of permalloy stripes. The data clearly exhibit larger frequency shifts for resonators with more stripes, reflecting an increase in the $G_{\mathrm{eff}}$. These results experimentally confirm theoretical predictions[39], demonstrating that $\Delta f_{\mathrm{BS}}$ scales positively with $G_{\mathrm{eff}}$, achieved by increasing the number of permalloy elements. The enhancement of $\Delta f_{\mathrm{BS}}$ at higher magnetic fields is attributed to an increase in virtual photon excitations induced by counter-rotating terms within the regime of $\omega_{\mathrm{p}}/\omega_{\mathrm{m}} \ll 1$[40].

To quantitatively validate the cooperative enhancement of the counter-rotating terms, we examined the scaling relationship between $\Delta f_{\mathrm{BS}}$ and the effective coupling strength. Under the lowest-order approximation, theory predicts the Bloch-Siegert shift to scale as $\Delta f_{\mathrm{BS}} \propto {G_{\mathrm{eff}}}^2/\omega_{\mathrm{p}}$. Figure 4b plots the measured $\Delta f_{\mathrm{BS}}$ against ${G_{\mathrm{eff}}}^2/f_{\mathrm{p}}$ at magnetic fields of 65 mT, 100 mT, and 140 mT. The linear dependencies observed confirm the theoretical scaling relationship $\Delta f_{\mathrm{BS}} \propto {G_{\mathrm{eff}}}^2/\omega_{\mathrm{p}}$. From linear fits to the experimental data, the slope coefficients $C$ are determined as be −0.617 ± 0.012 at 65 mT, −0.655 ± 0.006 at 100 mT, and −0.694 ± 0.005 at 140 mT. Despite variations in the intrinsic magnon and photon relaxation rate across systems, the Bloch-Siegert shift is consistently governed by coupling strength and resonance frequency. This observation supports the theoretical assertion that the population of virtual photons arising from counter-rotating terms remains robust and unaffected by relaxation processes[10].

## Discussion and Outlook

While our experiments were conducted at 10 K – where thermal fluctuations dominate over quantum fluctuations ($k_B T \gg \hbar\omega$) – the observed Bloch-Siegert shift and the suppression of self-interaction terms are governed by the fundamental Hamiltonian structure, which remains valid in the quantum regime. Therefore, our results serve as a classical emulation of Dicke physics. By cooling the system to millikelvin temperatures in dilution refrigerators, this platform can directly access ground-state quantum phenomena, such as vacuum squeezing and entanglement, utilizing the verified Hamiltonian parameters. To explore these quantum phenomena, the advanced measurement techniques such as quantum state tomography or noise measurement of magnon state will be essential[41–43].

Magnons have garnered considerable interest as promising media for next-generation information technologies, primarily owing to their tunability[44], nonlinearity[45], nonreciprocity[46], and intrinsic non-Hermitian dynamics[47]. Moreover, magnons can be efficiently converted into spin and charge currents through interfacial spin-orbit coupling. The spin currents generated through this mechanism can encode and transmit quantum information intrinsic to magnons[41,48–50], providing a promising route to explore entangled quantum states arising from counter-rotating terms in magnon-photon interactions. Accessing even stronger coupling regimes – such as deep-strong coupling – could be feasible by optimizing the coupling strength $g_s$ through advanced photonic architectures[29], by utilizing high-spin-density ferromagnetic materials, like CoFe thin films[51] or by harnessing the cooperative enhancement of separated ferromagnetic elements demonstrated in this work. Many proposed approaches for integrating magnonics into ultrastrong coupling phenomena can readily be implemented in our system by leveraging the Dicke cooperativity arising from separated ferromagnetic elements.

The suppressed, but still finite photon self-interaction term ($A^2$ term) in the Hopfield Hamiltonian – an unexpected feature in magnon-photon interactions – merits careful examination. In principle, the $A^2$ term arises from minimal coupling within electron–based-dipole interactions (see the details in Supplemental Information No.6)[20]. Therefore, the finite diamagnetic term observed in our system suggests a notable electronic contribution to the photon component in magnon polaritons. In contrast to previous studies on ultrastrongly coupled magnon polaritons in insulating YIG ferromagnets – where the insulating nature significantly reduces the magnitude of the $A^2$ shift[32] – the finite $A^2$ term identified in permalloy may originate from electron-mediated interactions, including electron-magnon, or electron-photon coupling, or potentially from geometric contributions associated with the specific gauge choice in lumped-element circuits. Given that itinerant electrons play the crucial role of orbital angular momentum in ferromagnets, quantitative evaluation of the $A^2$ term may provide essential insights into orbitronic phenomena. Although systematic material-controlled experiments and theoretical frameworks explicitly accounting for electron contributions in ferromagnets will be indispensable to comprehensively elucidate the precise mechanism underlying the $A^2$ term in magnon-polariton systems, pursuing this research direction promises not only fundamental insights into the $A^2$ term essential for achieving thermal equilibrium superradiant phase transition but also novel methods to probe orbital dynamics in ferromagnets, addressing longstanding challenges in spintronics.

## Methods

**Preparation of superconducting resonators.** The 150 nm-thick YBCO films, capped with a 40 nm-thick Au layer and deposited on sapphire substrates with dimensions of 6 mm × 10 mm, were sourced commercially. The YBCO chips consist of a single feedline coupled to five lumped-element resonators with varying resonance frequencies (see the Supplemental Information). These structures were patterned using electron-beam lithography, with the interdigital finger systematically modified across each resonator to tune their frequencies. The parameter of the resonator structure, as shown in Extended data Fig. 1a, is in the extended table 1, where $l_c$ is varied to tune the resonance frequency. The resonators are labelled R1 through R5, corresponding to increasing resonance frequencies. Subsequently, the YBCO and Au layers were etched using Ar ion milling to define the resonator patterns. The Au capping layer on the resonators was completely removed by an additional Ar ion milling step to suppress the eddy current loss in Au layer.

**Integration of Thin Ferromagnetic Stripes onto YBCO resonators.** A 30 nm-thick permalloy ($Ni_{80}Fe_{20}$) layer, patterned into stripes aligned with the inductor segments of the lumped-element resonators, was deposited on the YBCO resonators via electron-beam evaporation. To ensure electrical insulation, a 10 nm-thick $SiO_2$ buffer layer was first sputtered onto the surface. The full multilayer structure of the ferromagnetic stripes follows the stacking sequence: $SiO_2$ (10 nm) / MgO (2 nm) / Permalloy (30 nm) / Ti (3 nm) (see Fig. 2a) with the geometry of 15 μm × 400 μm.

**Transmission measurements.** The YBCO chips containing the resonator structures were placed in a low-temperature cryostat system, with RF cables connected to a vector network analyzer (VNA) (see Extended Data Fig. 1). Transmission spectra, $S_{12}$, were measured as a function of the external magnetic field, with the temperature held constant at 10 K. Here the transmission spectra are attenuated with 20 dB before and 6 dB after the YBCO chip, with a 18 dB amplification on the detection side. The external magnetic field was varied within a range of ±145 mT. The normalized transmission spectra, $|\Delta S_{12}|$, were calculated by dividing $S_{12}$ by the transmission spectrum measured at a magnetic field where no magnon-polariton peaks were observed, such as at 30 mT. This normalization effectively removes background contributions from environmental noise, isolating the relevant spectral features of the resonators.

**Modelling the notch-type resonators.** Using input-output formulism, the transmission spectra of the notch-type resonator structure can be expressed as following :

$$S_{21}(f) = \frac{\tilde{V}_2}{\tilde{V}_1} = ae^{i\beta}e^{-2\pi i f\tau}\left[1 - \frac{(Q/|Q_{\mathrm{ext}}|)e^{i\phi}}{1 + 2iQ(\frac{f}{f_{\mathrm{r}}} - 1)}\right], \quad (5)$$

$$Q = \left(\frac{1}{Q_{\mathrm{ext}}} + \frac{1}{Q_{\mathrm{int}}}\right)^{-1}, \quad (6)$$

where $ae^{i\beta}e^{-2\pi i f\tau}$ is the environment coefficient arising from the measurement system, excluding the resonator contribution. The term inside the square brackets represents the resonator response, incorporating an additional phase correction $e^{i\phi}$. The resonance frequency $f_{\mathrm{r}}$ is given by $1/\sqrt{LC}$. The $Q_{\mathrm{int}}$ and $Q_{\mathrm{ext}}$ correspond to the internal quality factor of the resonator and the external quality factor related to the capacitive coupling between the resonator and the feedline, respectively. The magnon polariton frequency $f_{\mathrm{mp}}$ can be estimated by Eq. (5) as a function of the external magnetic field.

The homogeneous AC magnetic field is excited on the inductor part of this type of lumped element superconducting resonator[54]. We perform finite-element simulations to map the field magnitude. The resulting profile reveals pronounced confinement of the photon mode within a few micrometers, which underpins an enhanced single-spin coupling strength.

**Kittel mode (*k* = 0 magnon mode).** The magnon mode with the uniform magnetization precession ($k = 0$) in thin ferromagnetic films with in-plane external magnetic field is expressed by the Kittel formula as follows:

$$f_{\mathrm{m}} = \frac{\gamma\mu_0}{2\pi}\sqrt{H(H + M_{\mathrm{eff}})}. \tag{7}$$

Here, $\gamma$, $\mu_0$, and $M_{\mathrm{eff}}$ are gyromagnetic ratio of ferromagnetic materials, the vacuum permeability, and the effective magnetization, respectively. In this mode, all the magnetic moment precesses in phase, resulting in a collective oscillation of the magnetization in thin ferromagnetic films without the spatial variation unlike $k \neq 0$ magnons. Considering the FEM result of excited AC magnetic field in this superconducting resonator[54], Kittel mode is excited in phase in the ferromagnetic stripes mounted on the meander structure of the superconducting resonator.

**Coupling Hamiltonian between one photon mode and collective magnon modes.** The total Hamiltonian describing the coupling between a photon mode and collective magnon modes is effectively captured by the gauge-invariant Hopfield model[26,55,56], which consistently incorporates self-interaction terms essential for satisfying the gauge principle:

$$\begin{aligned}\widehat{\mathcal{H}} = \hbar\omega_{\mathrm{p}}\hat{a}^\dagger\hat{a} + \hbar\sum_i \omega_{\mathrm{i}}\widehat{m}_i^\dagger\widehat{m}_i + \hbar\sum_i g_i(\hat{a}^\dagger + \hat{a})(\widehat{m}_i^\dagger + \widehat{m}_i) + \frac{\hbar}{\omega_{\mathrm{p}}}\sum_i {g_i}^2\left(\widehat{m}_i^\dagger + \widehat{m}_i\right)^2 \\ + \hbar D(\hat{a}^\dagger + \hat{a})^2.\end{aligned} \tag{8}$$

Here, $\omega_{\mathrm{p}}$ denotes the uncoupled photon resonance frequency, $\omega_{\mathrm{i}}$ are the uncoupled resonance frequencies of the $i$-th magnon mode, described by the Kittel formula. The term proportional to ${g_i}^2\big(\widehat{m}_i^\dagger + \widehat{m}_i\big)^2$ represents the magnetic self-interaction term of magnons arising from the dipole coupling within each ferromagnetic element, derived from the minimal coupling expression of magnon-photon interactions[26]. The term proportional to $(\hat{a}^\dagger + \hat{a})^2$ is introduced to explicitly for the diamagnetic self-energy (the $A^2$ term). While often negligible in pure spin systems, this term is included here to ensure a rigorous gauge-invariant description effectively evaluating the potential diamagnetic response from itinerant electrons in the metallic permalloy.

When each magnon mode oscillates coherently at frequency $\omega_{\mathrm{i}} = \omega_{\mathrm{m}}$ as illustrated in Fig. 3a, these local modes form a collective bright mode, represented by the operator[57]:

$$\widehat{M}_B = \frac{1}{G_{\mathrm{eff}}} \sum_i g_i \, \widehat{m}_i, \tag{9}$$

$$G_{\mathrm{eff}} = \sqrt{\sum_i |g_i|^2} = g_0 \sqrt{n}, \tag{10}$$

where $n$ is the total number of ferromagnetic elements, and $G_{\mathrm{eff}}$ is the effective coupling strength between the collective bright magnon mode and the photon mode. By substituting Eqs. (9) and (10) into Eq. (8), the interaction Hamiltonian (the third term in Eq. (8)) can be expressed as:

$$\mathcal{H}_{\mathrm{int}} = \hbar G'_{\mathrm{eff}} \Big( \hat{a}^\dagger \widehat{M}_{\mathrm{B}} + \hat{a} \widehat{M}_{\mathrm{B}}^\dagger + \hat{a}^\dagger \widehat{M}_{\mathrm{B}}^\dagger + \hat{a} \widehat{M}_{\mathrm{B}} \Big). \tag{11}$$

Here, we redefine the effective coupling as $G'_{\mathrm{eff}} = G_{\mathrm{eff}} \sqrt{\frac{\omega_{\mathrm{m}}}{\omega_{\mathrm{p}}}}$ to explicitly evaluate coupling parameters at low frequencies around $\omega_{\mathrm{m}}$.

The introduction of the collective bright mode requires a rigorous transformation of the magnon self-interaction term. We decompose the local magnon operators into the collective bright mode and a subspace of $n-1$ orthogonal dark modes $\widehat{M}_{\mathrm{D},k}$, expressed by:

$$\widehat{m}_i = \frac{g_i}{G_{\mathrm{eff}}} \widehat{M}_{\mathrm{B}} + \sum_{k=1}^{n-1} d_{ik} \widehat{M}_{\mathrm{D},k}. \tag{12}$$

Here, $d_{ik}$ represents the transformation coefficients for the $k$-th dark mode, satisfying the orthogonality condition $\sum_i g_i d_{ik} = 0$. Due to the spatial separation of the ferromagnetic elements, the self-interaction term sums locally ($\sum_i \widehat{m}_i{}^2$) rather than coherently ($(\sum_i \widehat{m}_i)^2$). Consequently, utilizing the orthogonality between the bright and dark modes, the collective expression for the self-interaction term transforms as follows:

$$\begin{aligned}\frac{\hbar}{\omega_{\mathrm{p}}}\sum_i g_i{}^2\left(\widehat{m}_i^\dagger+\widehat{m}_i\right)^2&=\frac{\hbar}{\omega_{\mathrm{p}}}\sum_i g_i{}^2\left[\frac{g_i}{G_{\mathrm{eff}}}\left(\widehat{M}_{\mathrm{B}}^\dagger+\widehat{M}_{\mathrm{B}}\right)+\sum_{k=1}^{N-1}d_{ik}(\widehat{M}_{D,k}^\dagger+\widehat{M}_{\mathrm{D},k})\right]^2\\&\approx\frac{\hbar g_0{}^2}{\omega_{\mathrm{p}}}\frac{1}{n}\sum_i\left(\widehat{M}_{\mathrm{B}}^\dagger+\widehat{M}_{\mathrm{B}}\right)^2=\frac{\hbar g_0{}^2}{\omega_{\mathrm{p}}}\left(\widehat{M}_{\mathrm{B}}^\dagger+\widehat{M}_{\mathrm{B}}\right)^2.\end{aligned}\tag{13}$$

Here, we define the effective self-interaction rate as $D_{\mathrm{m}}=\frac{g_0{}^2}{\omega_{\mathrm{p}}}$. Note that Eq. (11) demonstrates that $D_{\mathrm{m}}$ is determined by the single-element coupling $g_0$ and is not enhanced by the number of elements $n$, in stark contrast to the coupling strength $G_{\mathrm{eff}}\propto\sqrt{n}$. The dark modes are mathematically decoupled from the photon mode and the bright mode in the interaction Hamiltonian; thus, they do not contribute to the polariton formation and can be explicitly excluded from the effective Hamiltonian.

Thus the total Hamiltonian describing the interaction between the photon mode and the bright magnon mode is:

$$\begin{aligned}\mathcal{H}=\hbar\omega_{\mathrm{p}}\hat{a}^\dagger\hat{a}+\hbar\omega_{\mathrm{m}}\widehat{M}_{\mathrm{B}}^\dagger\widehat{M}_{\mathrm{B}}+\hbar G'_{\mathrm{eff}}\left(\hat{a}^\dagger\widehat{M}_{\mathrm{B}}+\hat{a}\widehat{M}_{\mathrm{B}}^\dagger+\hat{a}^\dagger\widehat{M}_{\mathrm{B}}^\dagger+\hat{a}\widehat{M}_{\mathrm{B}}\right)\\+\hbar D_{\mathrm{m}}\left(\widehat{M}_{\mathrm{B}}^\dagger+\widehat{M}_{\mathrm{B}}\right)^2+\hbar D(\hat{a}^\dagger+\hat{a})^2,\end{aligned}\tag{14}$$

Diagonalization of the Hamiltonian in Eq. (12) can be achieved through the Hopfield-Bogoliubov transformation, employing the operator vector $v^\dagger=[\hat{a}^\dagger\quad\widehat{M}_B^\dagger\quad\hat{a}\quad\widehat{M}_B]$, as follows[35]:

$$\mathcal{H}=\frac{1}{2}[\hat{a}^\dagger\quad\widehat{M}_{\mathrm{B}}^\dagger\quad\hat{a}\quad\widehat{M}_{\mathrm{B}}]\begin{bmatrix}\omega_{\mathrm{p}}+2D&G'_{\mathrm{eff}}&2D&G'_{\mathrm{eff}}\\G'_{\mathrm{eff}}&\omega_{\mathrm{m}}+2D_{\mathrm{m}}&G'_{\mathrm{eff}}&2D_{\mathrm{m}}\\2D&G'_{\mathrm{eff}}&\omega_{\mathrm{p}}+2D&G'_{\mathrm{eff}}\\G'_{\mathrm{eff}}&2D_{\mathrm{m}}&G'_{\mathrm{eff}}&\omega_{\mathrm{m}}+2D_{\mathrm{m}}\end{bmatrix}\begin{bmatrix}\hat{a}\\\widehat{M}_B\\\hat{a}^\dagger\\\widehat{M}_B^\dagger\end{bmatrix}.\tag{15}$$

Introducing polariton operators, defined as linear combinations of photon and bright magnon mode operators, $\hat{p}^\pm=\alpha^\pm\hat{a}+\beta^\pm\widehat{M}_{\mathrm{B}}+\gamma^\pm\hat{a}^\dagger+\delta^\pm\widehat{M}_{\mathrm{B}}^\dagger$, the Hamiltonian in Eq. (13) is diagonalized in terms of polariton frequencies as:

$$\mathcal{H}=\omega_{\mathrm{mp}}^+\hat{p}^{+\dagger}\hat{p}^++\omega_{\mathrm{mp}}^-\hat{p}^{-\dagger}\hat{p}^-,\tag{16}$$

which satisfies the eigenvalue equation $[\hat{p}^\pm,\mathcal{H}]=\omega_{\mathrm{mp}}^\pm\hat{p}^\pm$. Therefore, solving the eigenvalue problem of the Hopfield matrix $M$, explicitly given by:

$$M = \begin{bmatrix} \omega_{\rm p} + 2D & G'_{\rm eff} & 2D & G'_{\rm eff} \\ G'_{\rm eff} & \omega_{\rm m} + 2D_{\rm m} & G'_{\rm eff} & 2D_{\rm m} \\ 2D & G'_{\rm eff} & \omega_{\rm p} + 2D & G'_{\rm eff} \\ G'_{\rm eff} & 2D_{\rm m} & G'_{\rm eff} & \omega_{\rm m} + 2D_{\rm m} \end{bmatrix} \begin{bmatrix} 1 & 0 & 0 & 0 \\ 0 & 1 & 0 & 0 \\ 0 & 0 & -1 & 0 \\ 0 & 0 & 0 & -1 \end{bmatrix}. \quad (17)$$

yields the magnon-polariton frequencies $\omega_{\rm mp}^{\pm}$. The contributions of the counter-rotating and self-interaction terms such as $D$ and $D_{\rm m}$ can be selectively excluded, depending on the approximation required. Approximate analytical expressions for other polariton branches are provided in the Supplemental Information No. 4.

**Evaluation of spin coupling strength per Bohr magneton $g_{\rm s}$ in YBCO#1 and YBCO#2.**

To further characterize the YBCO resonator's performance, we evaluated the coupling strength per Bohr magneton ($g_{\rm s}$). The total number of electron spins in a permalloy stripe with the geometry of 15 μm × 400 μm – determined by using a saturation magnetization of permalloy; $\mu_0 M_{\rm s} = 1$ T[30] – is 1.55×10[13], leading to $g_{\rm s} \approx$ 28.4 Hz for YBCO#1 and $g_{\rm s} \approx$ 35.8 Hz for YBCO#2 from Eq. (1). They are obtained in resonators with one permalloy element in each YBCO chip by fitting the lower branch spectra with Dicke model. Here, we assume the same spin number with each permalloy element. Notably, these value values exceed the ~ 20 Hz reported for CPW resonators in previous studies[29,30], demonstrating a enhancement attributed to the optimized lumped-element resonator design, which features carefully engineered capacitor and inductor geometries.

**Critical coupling strength for superradiant phase transition.** We derive the critical coupling strength required to induce the superradiant phase transition in our system, where the magnon and photon self-interaction terms are effectively suppressed. Following the formalism established by Nataf and Ciuti[21], the phase critical point is identified by the onset of a soft mode where the energy of the lowest polariton branch vanishes. This corresponds to the condition $\det(M) = 0$. For the bilinear Hopfield Hamiltonian considered here, the determinant is given by:

$$\det(M) = \omega_{\rm p}\omega_{\rm m}\left(16DD_{\rm m} + 4D_{\rm m}\omega_{\rm p} + 4D\omega_{\rm m} - 4{G'_{\rm eff}}^2 + \omega_{\rm p}\omega_{\rm m}\right). \quad (18)$$

By substituting $D = \xi_{\rm p}\, {G'_{\rm eff}}^2/\omega_{\rm m}$ and $D_{\rm m} = \xi_{\rm m}\, {G'_{\rm eff}}^2\big/\omega_{\rm p}$ into Eq. (19), and solving Eq. (16) for the condition $\det(M) = 0$, we obtain the expression for the critical coupling strength:

$$G'_{\mathrm{eff,c}}{}^2 = \frac{\omega_\mathrm{p}\omega_\mathrm{m}}{8\xi_\mathrm{p}\xi_\mathrm{m}}\left\{(1-\xi_\mathrm{p}-\xi_\mathrm{m}) - \sqrt{(1-\xi_\mathrm{p}-\xi_\mathrm{m})^2 - 4\xi_\mathrm{p}\xi_\mathrm{m}}\right\}. \tag{19}$$

Here that $G'_{\mathrm{eff,c}}$ is defined to the effective coupling strength $G'_\mathrm{eff} = G_\mathrm{eff}\sqrt{\frac{\omega_\mathrm{m}}{\omega_\mathrm{p}}}$ . Thus, a finite real solution for $G'_{\mathrm{eff,c}}$ exists only when the condition $\xi_\mathrm{p} + \xi_\mathrm{m} < 1$ is satisfied. Conversely, in the regime where $\xi_\mathrm{p} + \xi_\mathrm{m} \geq 1$, the critical coupling becomes imaginary or undefined. This singularity corresponds to the standard no-go theorem scenario, which forbids the superradiant phase transition in the presence of unsuppressed self-interaction terms[20,21,26].

In the limit where the suppression factors $\xi_\mathrm{p}$ and $\xi_\mathrm{m}$ are sufficiently small, we can apply the Taylor expansion approximation:

$$\sqrt{(1-\xi_\mathrm{p}-\xi_\mathrm{m})^2 - 4\xi_\mathrm{p}\xi_\mathrm{m}} \approx 1-\xi_\mathrm{p}-\xi_\mathrm{m} - 2\frac{\xi_\mathrm{p}\xi_\mathrm{m}}{1-\xi_\mathrm{p}-\xi_\mathrm{m}}. \tag{20}$$

Substituting Eq. (21) into Eq. (20) yields the approximate expression:

$$G'_{\mathrm{eff,c}}{}^2 = \frac{\omega_\mathrm{p}\omega_\mathrm{m}}{4}\frac{1}{1-\xi_\mathrm{p}-\xi_\mathrm{m}}. \tag{21}$$

Finally, in the limit of vanishing self-interactions ($\xi_\mathrm{p}, \xi_\mathrm{m} \to 0$), we recover the standard Dicke model critical coupling: $G'_{\mathrm{eff,c}} = \sqrt{\omega_\mathrm{p}\omega_\mathrm{m}}/2$ [6,21]. This result confirms that our derived condition smoothly connects to the canonical critical coupling strength for the superradiant phase transition in the absence of the no-go theorem constraints.

**Analyzation of spectra with relaxation rate of magnons and photons.** To investigate the relaxation rate of magnons and photons, we analyze the magnon-polariton modes by explicitly including the respective relaxation rates for magnons ($\kappa_\mathrm{m}$) and photons ($\kappa_\mathrm{p}$). These relaxation rates of magnons and photons, which can be characterized via an input-output formalism[30], play a crucial role in evaluating cooperativity and decoherence time of the system and discussing non-Hermitian physics[47]. We adopt a simplified Tavis–Cummings model that neglects both counter-rotating terms and the diamagnetic term, resulting in the effective Hamiltonian considering $\kappa_{\rm m}$ and $\kappa_{\rm p}$:

$$\mathcal{H} = \hbar(\omega_{\rm p} - i\kappa_{\rm p})\hat{a}^\dagger\hat{a} + \hbar(\omega_{\rm m} - i\kappa_{\rm m})\widehat{M}_{\rm B}^\dagger\widehat{M}_{\rm B} + \hbar G_{\rm eff}\left(\hat{a}^\dagger\widehat{M}_{\rm B} + \hat{a}\widehat{M}_{\rm B}^\dagger\right). \tag{22}$$

The corresponding Heisenberg equations of motion for the annihilation operators $\hat{a}$ (photon mode) and $\widehat{M}$ (collective magnon mode) are:

$$\frac{d\hat{a}}{dt} = -i\omega_{\rm p}\hat{a} - G_{\rm eff}\widehat{M} - \kappa_{\rm p}\hat{a}, \tag{23}$$

$$\frac{d\widehat{M}}{dt} = -i\omega_{\rm m}\widehat{M} - G_{\rm eff}\hat{a} - \kappa_{\rm m}\widehat{M}. \tag{24}$$

Assuming harmonic solutions $\hat{a}(t) = a_0 e^{-i\omega t}$, $\widehat{M}(t) = m_0 e^{-i\omega t}$, we derive the matrix equation:

$$\begin{pmatrix} i(\omega_{\rm p} - \omega) + \kappa_{\rm p} & G_{\rm eff} \\ G_{\rm eff} & i(\omega_{\rm m} - \omega) + \kappa_{\rm m} \end{pmatrix} \begin{pmatrix} a_0 \\ b_0 \end{pmatrix} = 0. \tag{25}$$

Solving for eigenfrequencies gives:

$$\tilde{\omega}_\pm = \frac{\omega_{\rm m} + \omega_{\rm p} - i(\kappa_{\rm p} + \kappa_{\rm m})}{2} \pm \frac{1}{2}\sqrt{[\omega_{\rm p} - \omega_{\rm m} - i(\kappa_{\rm m} - \kappa_{\rm p})^2]^2 + 4{G_{\rm eff}}^2}. \tag{26}$$

The real parts of $\tilde{\omega}_\pm$ represent resonance frequencies, while the imaginary parts correspond to relaxation rates of the magnon-photon polaritons.

In the limit $f_{\rm i} \gg \kappa_{\rm i}$ for uncoupled magnon and photon modes, the resonance frequencies $\omega_\pm$ and the relaxation rates $\kappa_\pm$ reduce to:

$$\omega_\pm = \frac{\omega_{\rm m} + \omega_{\rm p}}{2} \pm \frac{1}{2}\sqrt{[\omega_{\rm p} - \omega_{\rm m}]^2 + 4{G_{\rm eff}}^2}, \tag{27}$$

$$\kappa_\pm = \frac{\kappa_{\rm m} + \kappa_{\rm p}}{2} \pm \frac{\kappa_{\rm p} - \kappa_{\rm m}}{2}\frac{\omega_{\rm p} - \omega_{\rm m}}{\sqrt{[\omega_{\rm p} - \omega_{\rm m}]^2 + 4{G_{\rm eff}}^2}}. \tag{28}$$

Fitting experimental data for resonance frequencies and half-width at half-maximum (HWHM) enables the extraction of key parameters $G_{\rm eff}$, $\kappa_{\rm p}$, and $\kappa_{\rm m}$, thus identifying the coupling regime such as weak coupling, strong coupling or Purcell regime[28]. Extended figure 2 presents fitting results for a resonator coupled to a single permalloy stripe in YBCO#2, yielding $G_{\rm eff}/2\pi$ ≈ 129 ± 1.4 MHz, $\kappa_{\rm p}/2\pi$ ≈ 0.53 ± 0.03 MHz, and $\kappa_{\rm m}/2\pi$ ≈ 461 ± 11 MHz, placing these magnon polaritons in the Purcell regime ($\kappa_{\rm p} < G < \kappa_{\rm m}$)[28]. Notably, the broader linewidth observed for the magnon-like branch compared with the expectation from Eq. (24) indicates a Purcell-enhanced relaxation of the photons driven by the large magnon relaxation rate in this system. Assuming $\kappa_{\mathrm{m}}$ and $\kappa_{\mathrm{p}}$ remain unchanged when multiple magnon modes couple to a single photon mode, the cooperativity $C = {G_{\mathrm{eff}}}^2/\kappa_{\mathrm{m}}\kappa_{\mathrm{p}}$ increases from 67.4 for one permalloy stripe to 1860 for 26 permalloy stripes. The highest cooperativity in our system surpasses previously reported values for magnon polaritons in thin ferromagnetic films coupled to superconducting resonator[30,58].

## Data availability

The data that support the findings of this work are available from the corresponding authors upon reasonable request.

## Acknowledgment

This work was supported in part by the Japan Society for the Promotion of Science (JSPS) Research Fellow Program (grant no. 22KJ1956), JSPS Overseas Challenge Program for Young Researchers (grant no. 202280145), Grant-in-Aid for Scientific Research (B) (grant no. 21H01798), and the Spintronics Research Network of Japan (Spin-RNJ). SY thanks T. Hioki and J. C. Gartside for the fruitful discussions and comments on the manuscript. MM, MA, and HH acknowledge financial support by the Deutsche Forschungsgemeinschaft (DFG, German Research Foundation) via Germany's Excellence Strategy EXC-2111- 390814868 and the Transregio "Constrained Quantum Matter" (TRR 360, Project-ID 492547816). This research is part of the Munich Quantum Valley, which is supported by the Bavarian State Government with funds from the Hightech Agenda Bayern Plus.

## Supplemental Information

No. 1 Evaluation of bare YBCO superconducting resonators

No. 2 Magnetic field and temperature dependence of YBCO resonator with lumped-element structure

No. 3 Surface image of YBCO resonators by AFM

No. 4 Derivation of eigenfrequencies with the different type of Hamiltonians for magnon polaritons

No. 5 Cavity-mediated coupling between permalloy elements

No. 6 Origin of $A^2$ term in electron-photon interactions

No. 7 Comparison with spectra under unsuppressed self-interaction conditions

## Figure captions

**Figure 1 | Design principle to achieve the suppression of magnon self-interaction term.**

**(a, b)** Schematic illustration of the coupling between the uniform photon mode and magnon modes excited in (a) single and (b) multiple ferromagnetic elements. Magnon modes excited in-phase in multiple ferromagnets collectively form magnon bright mode, which allows the enhancement of effective coupling strength scaling with $\sqrt{n}$, where $n$ is the number of the ferromagnetic elements.

**Figure 2 | On-chip magnon polaritons in the ultrastrong coupling regime and direct observation of the Bloch-Siegert shift.**

**(a)** Schematic illustration of the lumped-element superconducting resonator with 30-nm-thick permalloy (Py) stripes, with the inset depicting the layer structure of the ferromagnetic elements.

**(b)** The magnitude of the transmission spectrum $|S_{21}|$ of the bare resonator measured at 10 K and zero magnetic field. The red dashed curve is a fit via Eq. (5), see Methods.

**(c)** Photon absorption of the superconducting resonator incorporating multiple permalloy elements as a function of the applied magnetic field. Peaks shift clearly with varying magnetic field. The red dotted curves trace the peak positions extracted from the transmission spectra.

**(d-f)** Enlarged transmission spectra at low magnetic fields: (d) 0 mT, (e) 7 mT and (f) 14 mT, fitted by Eq. (5). These spectra demonstrate the subtle peak shifts attributed to magnon-photon coupling.

**(g)** Normalized transmission spectra $|\Delta S_{21}|$ as a function of applied magnetic field for the resonator with multiple permalloy stripes in YBCO#1 chip. Red dashed curves represent fits based on the polariton frequencies of Eq. (2). Blue and black dashed lines indicate the uncoupled photon and magnon modes, respectively. Inset shows extracted effective coupling strength $G_{\mathrm{eff}}$ and the normalized coupling ratio $\eta$, confirming the ultrastrong coupling regime ($\eta > 0.1$).

**(h, i)** Experimental spectra compared with simulations obtained by excluding the counter-rotating terms (CRTs) (purple dashed lines) and the diamagnetic term (orange dashed lines), for the upper **(d)** and lower polariton branches. Deviations induced by neglecting counter-rotating terms directly manifest as the Bloch-Siegert shift ($\Delta f_{BS}$).

**Figure 3 | Dicke cooperativity of collective magnon modes**

**(a-d)** Transmission spectra of resonators (b) 1, (c) 8, (d) 16 and (e) 26 permalloy stripes in YBCO#2 as a function of the applied magnetic field.

**(e)** Normalized coupling strength $\varepsilon$ as a function of $\sqrt{n}$, where $n$ is the number of permalloy stripes, for both YBCO#1 and YBCO#2.

**Figure 4 | Control of Bloch-Siegert shift via the effective coupling strength of magnon polaritons.**

**(a)** Bloch–Siegert shift $\Delta f_{BS}$ as a function of the applied magnetic field for resonators with different numbers of permalloy stripes. The inset identifies the YBCO chip index and the corresponding stripe count.

**(b)** Bloch–Siegert shift as a function of the square of the normalized coupling strength $\varepsilon$ for thin permalloy stripes, measured at applied magnetic fields of 65, 100, and 140 mT.

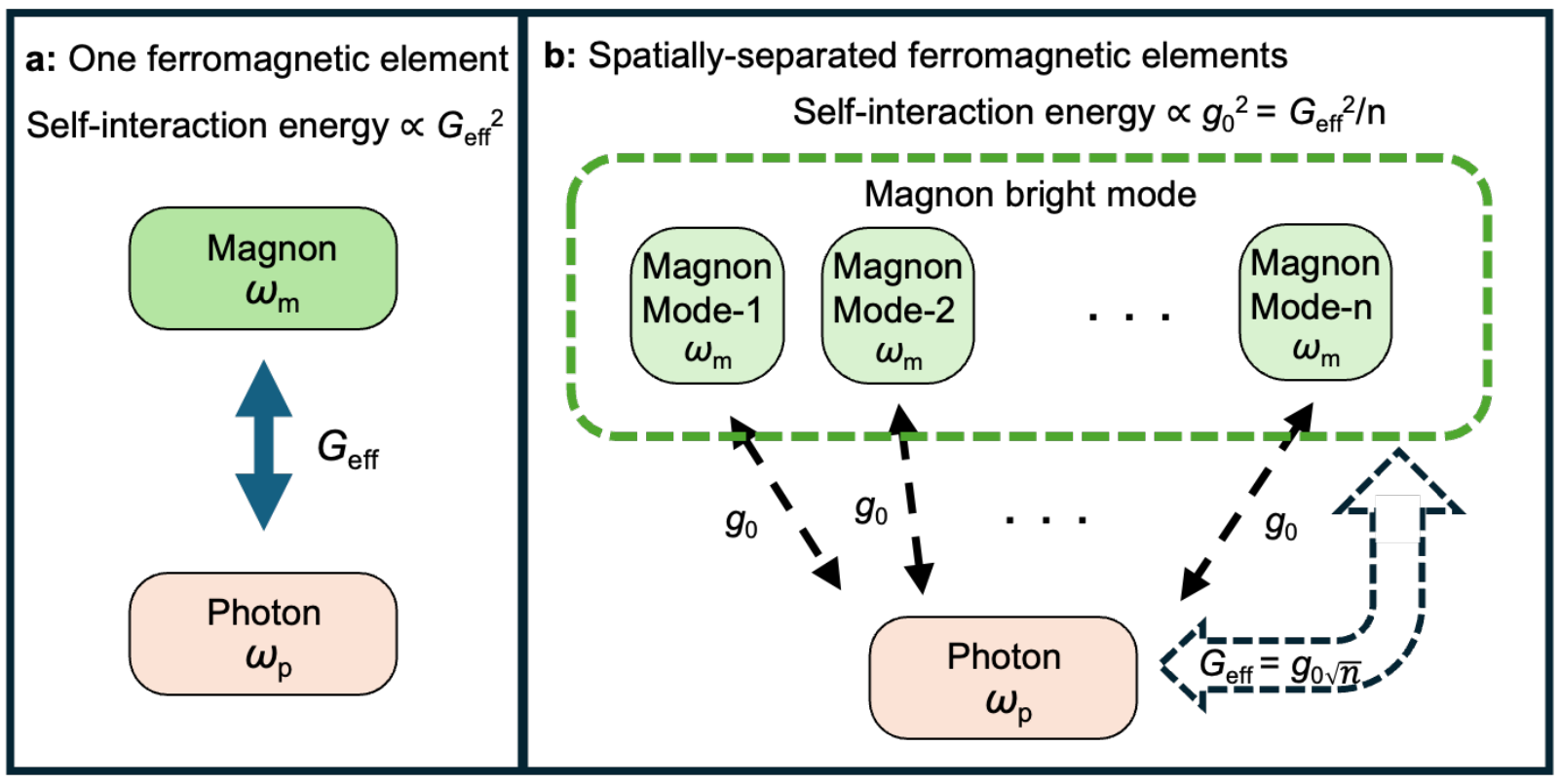

Fig. 1 Yoshii *et al.*

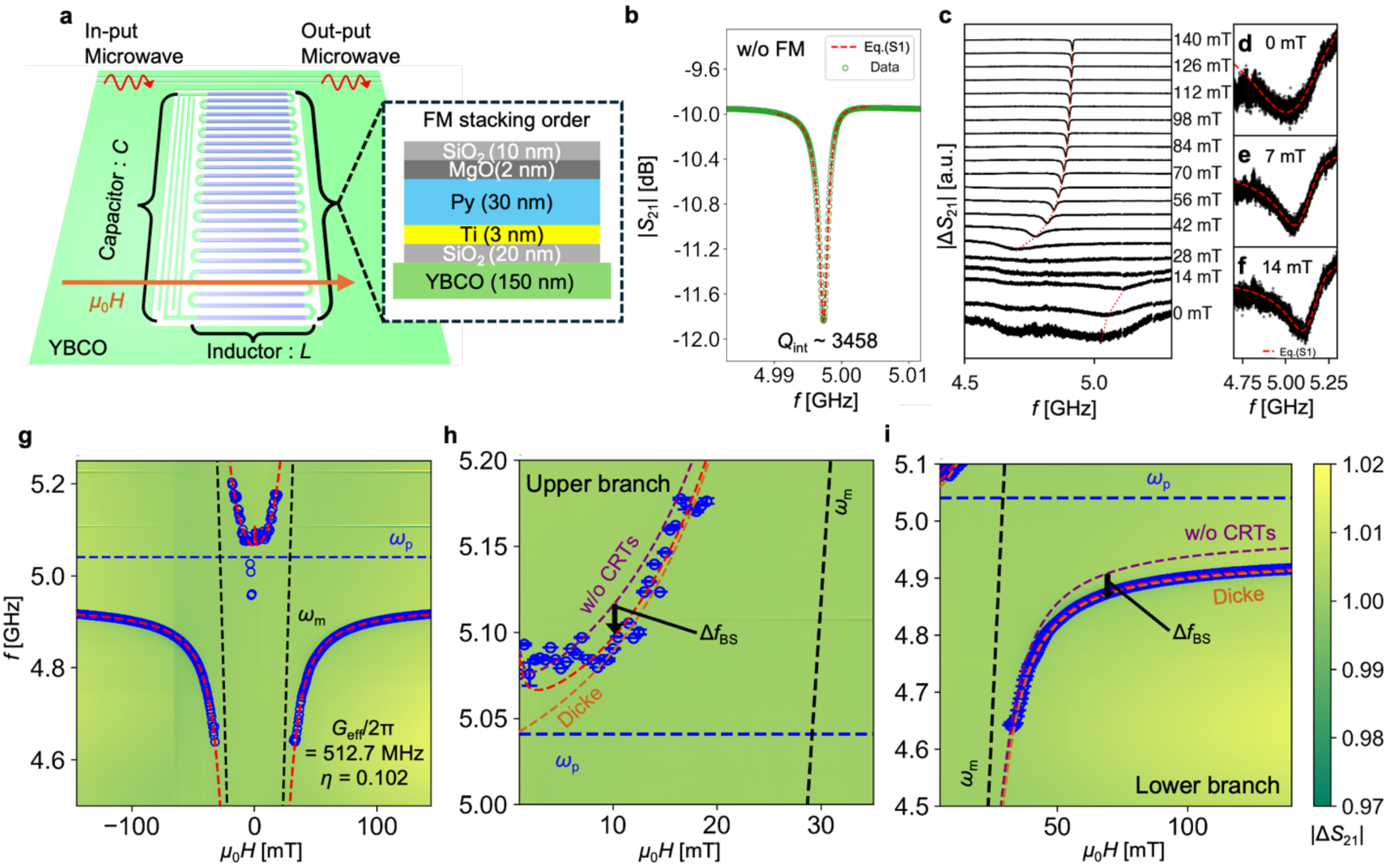

Fig. 2 Yoshii *et al.*

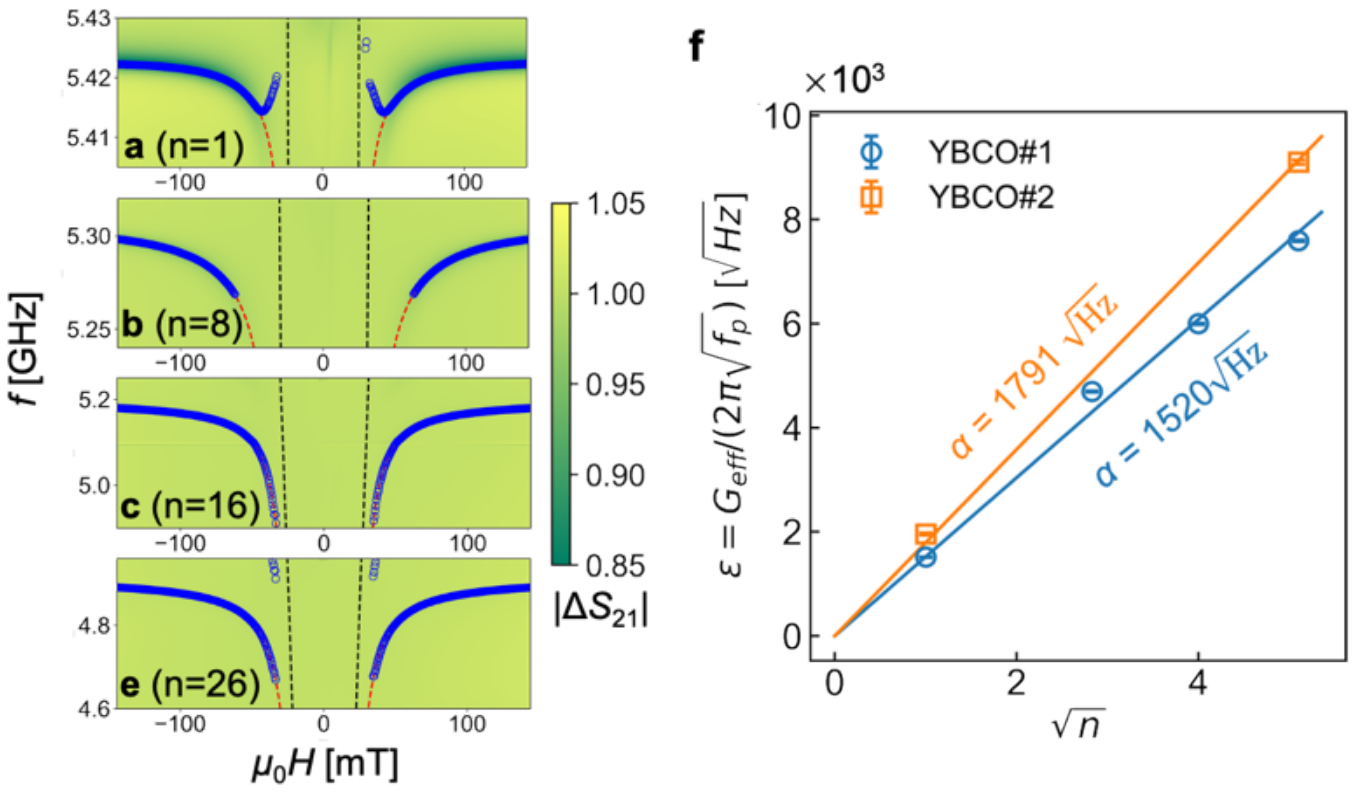

Fig. 3 Yoshii *et al.*

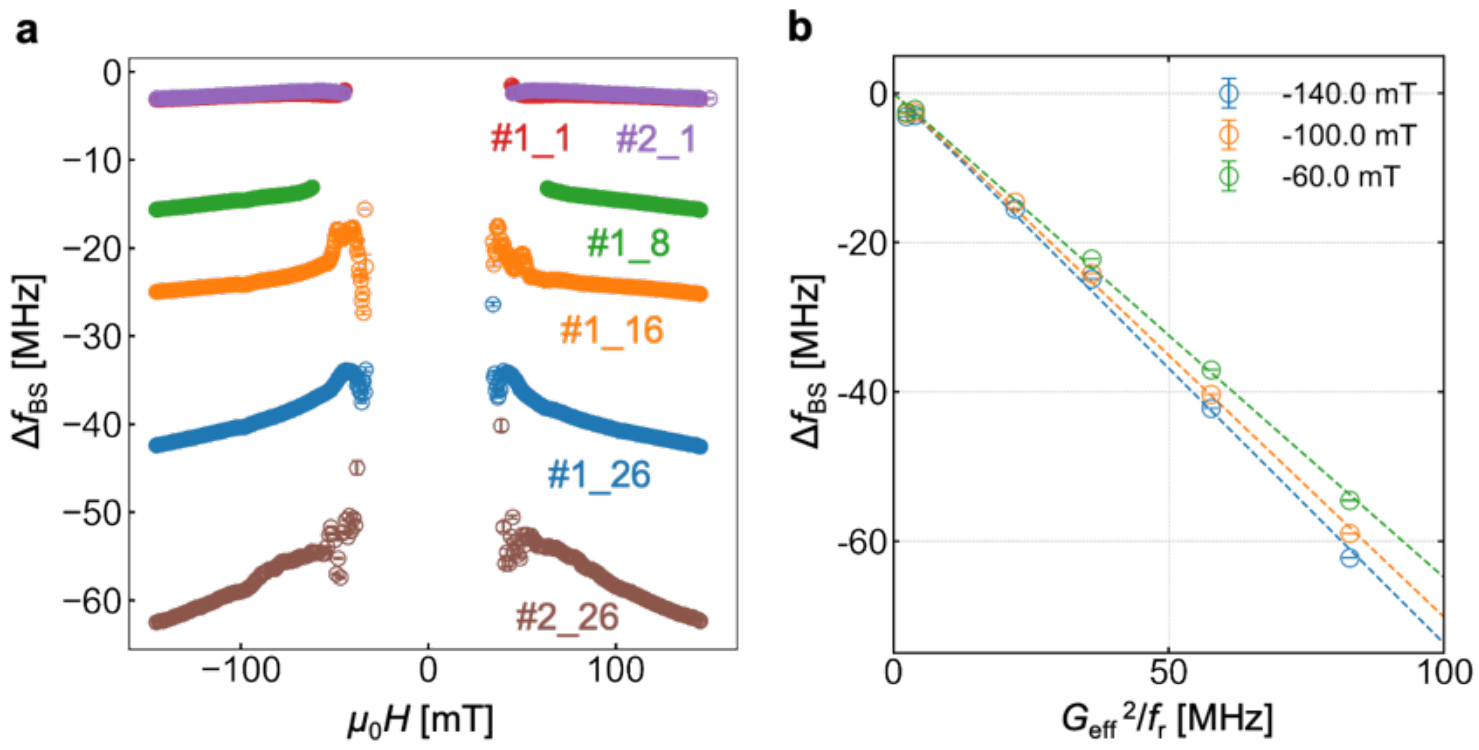

Fig. 4 Yoshii *et al.*

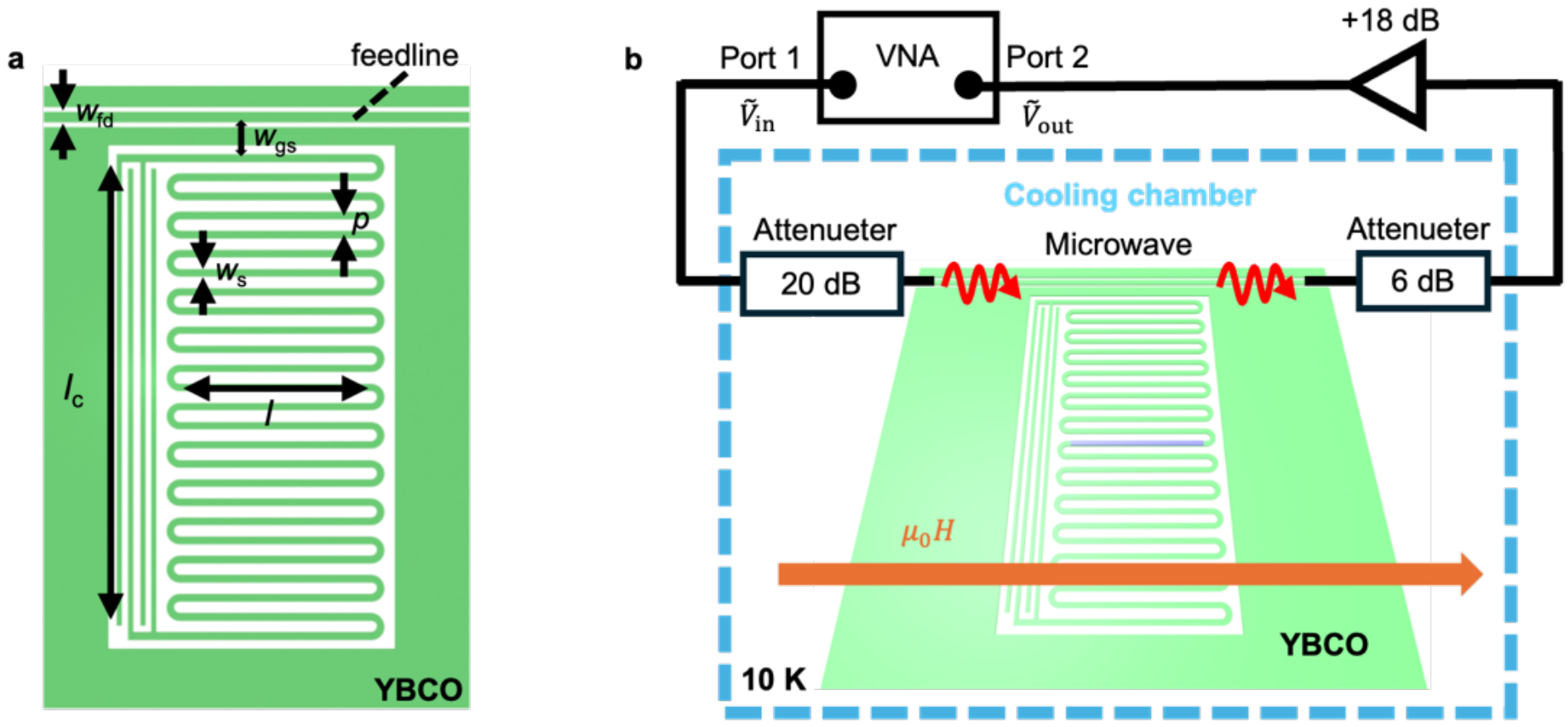

Extended Data Fig. 1

| Resonator parameter | |
|---|---|
| $w_{fd}$ | 20 µm |
| $w_s$ | 15 µm |
| $w_{gs}$ | 42 µm (YBCO#1)<br>72 µm(YBCO#2) |
| $p$ | 25 µm |
| $l$ | 400 µm |

Extended Table. 1

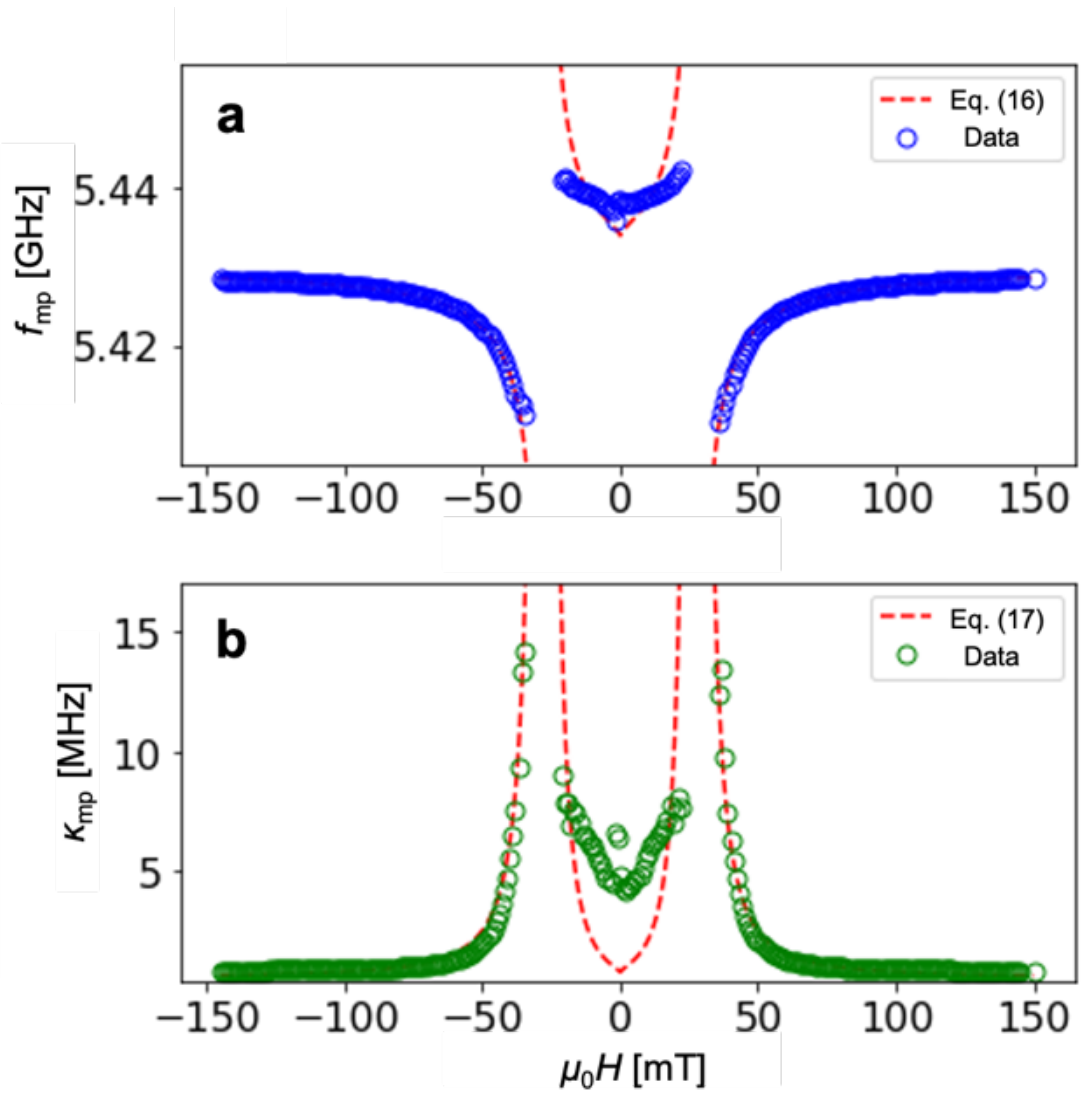

Extended Data Fig. 2